\documentclass[twocolumn,showpacs,jcp,superscriptaddress]{revtex4}
\usepackage{graphicx}
\usepackage[usenames]{color}
\usepackage{amssymb}
\usepackage{amsmath}
\usepackage{amsfonts}
\usepackage{mathrsfs}

\renewcommand{\epsilon}{\varepsilon}
\newcommand{\figurewidth}{0.46\textwidth}
\newcommand{\narrowfigurewidth}{0.25\textwidth}

\begin{document}
\title{Polymer translocation into a fluidic channel through a nanopore}

\author{Kaifu Luo}
\altaffiliation[]{Author to whom the correspondence should be addressed}
\email{klu@ustc.edu.cn}
\affiliation{CAS Key Laboratory of Soft Matter Chemistry, Department of Polymer
Science and Engineering, University of Science and Technology of China, Hefei, Anhui
Province 230026, P. R. China}

\author{Ralf Metzler}
\email{metz@ph.tum.de}
\affiliation{Physics Department, Technical University of Munich,
D-85748 Garching, Germany}

\date{\today}

\begin{abstract}

Using two dimensional Langevin dynamics simulations, we investigate the dynamics of polymer translocation into a fluidic channel 
with diameter $R$ through a nanopore under a driving force $F$.
Due to the crowding effect induced by the partially translocated monomers, the translocation dynamics is significantly altered in 
comparison to an unconfined environment, namely, we observe a nonuniversal dependence of the translocation time $\tau$ on the 
chain length $N$.
$\tau$ initially decreases rapidly and then saturates with increasing $R$, and a dependence of the scaling exponent $\alpha$ of 
$\tau$ with $N$ on the channel width $R$ is observed.
The otherwise inverse linear scaling of $\tau$ with $F$ breaks down and we observe a minimum of $\alpha$ as a function of $F$.
These behaviors are interpreted in terms of the waiting time of an individual segment passing
through the pore during translocation.

\end{abstract}


\maketitle

\section{Introduction}

Polymer translocation through a nanopore has attracted broad interest
because it is of fundamental relevance in polymer physics and is also related to
many biological processes, such as DNA and RNA translocation across nuclear
pores, protein transport through membrane channels, or viruses injecting their DNA
into a cell.
Due to its potentially revolutionary technological
applications~\cite{Kasianowicz,Meller03}, including rapid DNA sequencing, gene
therapy and controlled drug delivery, a considerable number of recent
experimental~\cite{Meller00,Meller01,Meller02,Akeson,Meller07,Bashir,Sauer,Mathe,Henrickson,
Kasianowicz2,Kasianowicz3,Kasianowicz4,Robertson,
Li01,Li03,Li05,Keyser1,Keyser2,Dekker,Trepagnier,Storm03,Storm052,Storm05} and
theoretical~\cite{Storm05,Simon,Sung,Park,diMarzio,Muthukumar99,MuthuKumar03,Kong,Lubensky,
Kafri,Slonkina,Matysiak,Ambj,Metzler,Ambj2,Ambj3,Baumg,Chuang,Kantor,
Dubbeldam1,Dubbeldam2,Milchev,Luo1,Luo2,Huopaniemi1,Huopaniemi2,Luo3,Luo4,Luo6,Luo7,Luo8,Panja,
Aniket,Slater,Sakaue,Chern,Loebl,Randel,Lansac,Farkas,Tian,Lu,Liao,Zandi,Tsuchiya,Kotsev,Bockelmann}
studies have been devoted to this subject.

The average translocation time $\tau$ as a function of the chain length $N$ is an important
measure of the underlying dynamics. Standard equilibrium Kramers analysis
\cite{Kramers} of diffusion across an entropic barrier yields $\tau \sim N^2$
for unbiased translocation and $\tau \sim N$ for driven translocation (assuming
friction to be independent of $N$) \cite{Sung,Muthukumar99}.
However, the quadratic scaling
behavior for unbiased translocation cannot be correct for a self-avoiding
polymer \cite{Chuang} because the translocation time would be shorter than the Rouse
equilibration time of a self-avoiding polymer, $\tau_R \sim N^{1+2\nu}$, where
the Flory exponent $\nu=0.588$ in 3D and $\nu_{2D}=0.75$ in 2D \cite{deGennes,Rubinstein}.
This renders the concept of equilibrium entropy and the ensuing entropic
barrier inappropriate for translocation dynamics. Chuang \textit{et al}.
\cite{Chuang} studied the translocation for both phantom and self-avoiding
polymers by numerical simulations with Rouse dynamics for a 2D
lattice model and showed that for large $N$, $\tau \sim N^{1+2\nu}$, the same scaling
behavior as the equilibration time but with a much larger prefactor.
This result was recently corroborated by extensive numerical simulations based
on the Fluctuating Bond (FB) \cite{Luo1} and Langevin Dynamics (LD) models with
the bead-spring approach \cite{Huopaniemi1,Liao}.

For driven translocation, Kantor and Kardar \cite{Kantor} demonstrated
that the assumption of equilibrium in polymer dynamics breaks down even more easily
and provided a lower bound $\tau \sim N^{1+\nu}$ for the translocation time by
comparison to the unimpeded motion of the polymer. Using FB \cite{Luo2} and LD
\cite{Huopaniemi1,Luo3} models, a crossover from $\tau \sim N^{2\nu}$ for
relatively short polymers to $\tau \sim N^{1+\nu}$ for longer chains was found
in 2D. In 3D, we find that for faster translocation processes $\tau \sim
N^{1.37}$ \cite{Luo6,Luo7}, while it crosses over to $\tau \sim N^{1+\nu}$ for
slower translocation, such as under weak driving forces and/or of high
friction constants \cite{Luo8}.
Moreover, using linear response theory with memory effects and some non-trivial assumptions,
Vocks {\em et. al.} \cite{Panja} came up with an alternative estimate $\tau \sim
N^{\frac{1+2\nu}{1+\nu}}$ for 3D, which means $\alpha=1.37$ in 3D.
Their $\alpha$ in 3D is consistent with our numerical data for fast translocation, but 
fails to capture the scaling exponent for slow translocation.

However, above physical pictures are based on translocation into an unconfined
\textit{trans} side. Even for the case of an unconfined \textit{trans} side, the
translocated chain is highly compressed during the translocation process under
fast translocation conditions \cite{Luo8}, and thus even more severe
nonequilibrium effects are expected under confinement. Quite little attention
has been paid to the dynamics of translocation into confined environments. We
here quantify the effects of the large entropic penalty on the confined chain
and show that it significantly affects the translocation dynamics. In particular we find
folded configurations of the confined chain on the \textit{trans} side right after
completion of the translocation process. Moreover we show that the force
dependence of the translocation time significantly changes the $1/F$
dependence for free translocation, and that the almost symmetric form of
the waiting time distribution turns over to a rapid increase throughout
the translocation process.
The paper is organized as follows. In section II, we briefly describe our model
and the simulation technique. In section III, we present our results. Finally,
we conclude in section IV.

\section{Model and methods} \label{chap-model}

In our numerical simulations, the polymer chains are modeled as bead-spring
chains of Lennard-Jones (LJ) particles with the Finite Extension Nonlinear
Elastic (FENE) potential. Excluded volume interaction between beads is
modeled by a short range repulsive LJ potential: $U_{LJ} (r)=4\epsilon
[{(\frac{\sigma}{r})}^{12}-{(\frac{\sigma} {r})}^6]+\epsilon$ for $r\le
2^{1/6}\sigma$ and 0 for $r>2^{1/6}\sigma$. Here, $\sigma$ is the diameter of a
bead, and $\epsilon$ is the depth of the potential. The connectivity between
neighboring beads is modeled as a FENE spring with $U_{FENE}
(r)=-\frac{1}{2}kR_0^2\ln(1-r^2/R_0^2)$, where $r$ is the distance between
consecutive beads, $k$ is the spring constant and $R_0$ is the maximum
allowed separation between connected beads.

\begin{figure}
  \includegraphics*[width=\narrowfigurewidth]{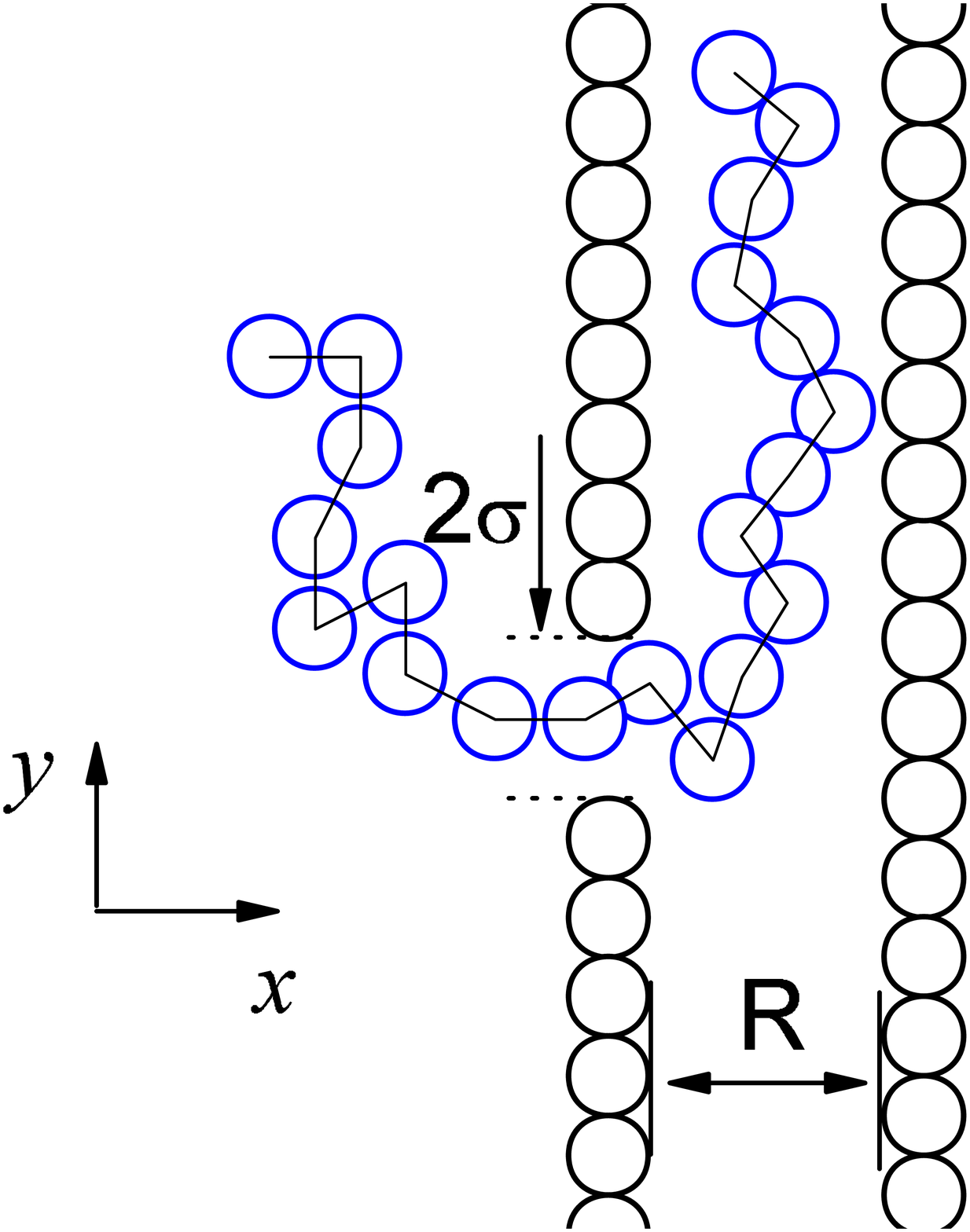}
\caption{(Color online) A schematic representation of polymer translocation through a pore
into a 2D
confined environment under an external driving force $F$ across the pore. The
pore width is $2\sigma$.
        }
 \label{Fig1}
\end{figure}

We consider a geometry as shown in Fig. \ref{Fig1}, where two strips with
separation $R$ are formed by stationary particles within a distance $\sigma$
from each other. One strip (``wall'') has a pore of diameter $2\sigma$.
Between all bead-wall particle pairs, there exists the same short range
repulsive LJ interaction as described above.
In the Langevin dynamics simulation, each bead is subjected to conservative,
frictional, and random forces, respectively, with~\cite{Allen} $m{\bf \ddot
{r}}_i =-{\bf \nabla}({U}_{LJ}+{U}_{FENE})+{\bf F}_{\textrm{ext}}-\xi {\bf
v}_i+{\bf F}_i^R$. Here $m$ is the bead's mass, $\xi$ is the friction
coefficient, ${\bf v}_i$ is the bead's velocity, and ${\bf F}_i^R$ is the
random force which satisfies the fluctuation-dissipation theorem.
The external force is expressed as ${\bf F}_{\textrm{ext}}=F\hat{x}$, where $F$
is the external force strength exerted on the beads in the pore, and
$\hat{x}$ is a unit vector in the direction along the pore axis.

In the present work, the LJ parameters $\epsilon$, $\sigma$, and $m$
fix the system energy, length and mass units respectively, leading to the
corresponding time scale $t_{LJ}=(m\sigma^2/\epsilon)^{1/2}$ and
force scale $\epsilon/\sigma$, which are of the order of ps and pN, respectively.
The dimensionless parameters in the model are then chosen to be
$R_0=2$, $k=7$, $\xi=0.7$, and $F=0.5\ldots15$.

In our model, each bead corresponds to a Kuhn length (twice of the persistence length) of a
polymer. For a single-stranded DNA (ssDNA), the persistence length of the ssDNA is sequence and
solvent dependent and varies in a wide range, to our knowledge, usually from about 1 to 4 nm.
We assume the value of $\sigma \sim 2.8$ nm for a ssDNA containing approximately four nucleotide bases.
The average mass of a base in DNA is about 312 amu, so the bead mass $m \approx 1248$ amu. We set
$k_{B}T=1.2\epsilon$, which means that the interaction strength $\epsilon$ is $3.39 \times 10^{-21}$ J 
at actual temperature 295 K. This leads to a time scale of 69.2 ps and a force scale of 1.2 pN.
Each base (nucleotide) is estimated to have an effective charge of 0.094e from Ref. \cite{Sauer}, leading 
to an effective charge of a bead with four bases of 0.376e.
Thus, the voltage across the pore is between 28.1 and 843 mV for varying $F$ from 0.5 to 15, within the 
range of experimental parameters \cite{Kasianowicz,Meller00,Meller01,Meller02,Meller03}.

The Langevin equation is then integrated in time by a method described by Ermak and Buckholz \cite{Ermak}. 
Initially, the first monomer of the chain is placed in the entrance of the pore, while the remaining 
monomers are under thermal collisions described by the Langevin thermostat to obtain an equilibrium 
configuration. Typically, we average our data over 1000 independent runs.

In Fig. \ref{Fig1}, we assume that both ends of the channel are open. If that were not the case, such 
as for virus capsids, on the entering of the chain the fluid in the confined region would necessarily 
have to leave. To address such a problem it is therefore relevant to explicitly include the solvent in 
the model, which is beyond the scope of the present study.

\section{Results and discussion} \label{chap-results}

Consider a polymer, such as DNA, confined to a nanochannel of width $R$ with $R$ being less than the 
radius of gyration of the molecule.
The response of the polymer to confinement is primarily dictated by the relative value of $R$ with respect 
to the chain persistence length. Depending on whether $R$ is larger (de Gennes regime) or smaller (Odijk regime) 
than the chain persistence length, different scaling behaviors of the longitudinal size of the chain, 
$ R_{\parallel}$, as a function of $R$ were predicted in the pioneering theoretical studies by de Gennes \cite{deGennes} 
and Odijk \cite{Odijk}, respectively.
In the Odijk regime, the physics is dominated not by excluded volume but by the interplay of confinement and 
intrinsic polymer elasticity.
In this work, we only consider the de Gennes regime, where the blob picture \cite{deGennes} is valid. To consider 
the Odijk regime, we would need to take into account the chain stiffness in the model. While this is certainly 
interesting we here focus on the polymeric aspects in the flexible chain limit.

According to the blob picture, for a polymer confined between two strips
embedded in 2D the chain will extend along the channel forming blobs of size $R$.
Each blob contains $g=(R/\sigma)^{1/\nu_{2D}}$ monomers, where $\nu_{2D}$ is the Flory
exponent in 2D \cite{Rubinstein,deGennes}, and the number of blobs is
$n_b=N/g=N(\sigma/R)^{1/\nu_{2D}}$. Then, the blob picture predicts the
longitudinal size of the chain to be
$ R_{\parallel}\sim n_bR\sim N\sigma(\frac{\sigma}{R})^{1/\nu_{2D}-1}\sim NR^{-1/3}$
\cite{Rubinstein,deGennes}.
The longitudinal relaxation time $\tau_{\parallel}$ is defined as the time
needed by a polymer to move a distance of the order of its longitudinal size,
$R_{\parallel}$.
Thus, $\tau_{\parallel}$ scales as $\tau_{\parallel} \sim \frac
{R_{\parallel}^2} {\widetilde{D}} \sim N^{3}R^{-2/3}$, with ${\widetilde{D}}=1/N$
being the diffusion constant.
The free energy cost in units of $k_BT$ is $\mathcal{F}=N(\sigma/R)^{1/\nu_{2D}}$.

For polymer translocation into confined environments, a driving force is
necessary to overcome the entropic repulsion $f(R)$ exerted by already
translocated monomers. Due to the highly non-equilibrium property
of the translocation process, it is difficult to estimate the resisting force
$f(R)$. If we assume that, for slow translocation processes, the resisting force
$f(R)$ scales as $f(R)=CR^{\gamma}$, with $C$ and $\gamma$ being the
associated prefactor and the scaling exponent, respectively;
then, under an external driving force
$F$ in the pore the translocation time $\tau$ can be written as $\tau \sim
\frac{N^{\alpha}}{F-f(R)} \sim \frac{N^{\alpha}}{F(1-CR^{\gamma}/F)}$ with
$\alpha$ being the scaling exponent of $\tau$ with chain length $N$. Due to
$\tau_{\infty} \sim \frac{N^{\alpha}}{F}$ for an unconfined system with $R \sim
\infty$, we have $1-\frac {\tau_{\infty}}{\tau} \sim CR^{\gamma}/F$. Based on
this relationship, we examine the dependence of $\tau$ on $R$.

\subsection{Translocation time as a function of the driving force}

\begin{figure}
\includegraphics*[width=\figurewidth]{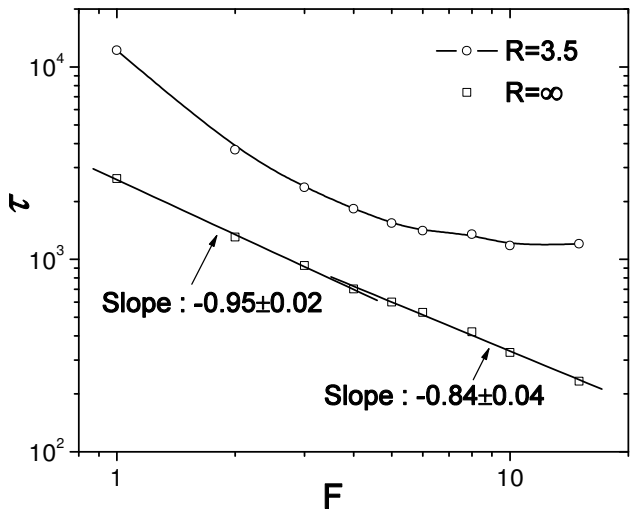}
\caption{Translocation time $\tau$ as a function of $F$ for $R=3.5$ and $R=\infty$ in 2D.
The chain length is $N=128$.
        }
\label{Fig2}
\end{figure}

\begin{figure}
\includegraphics*[width=\figurewidth]{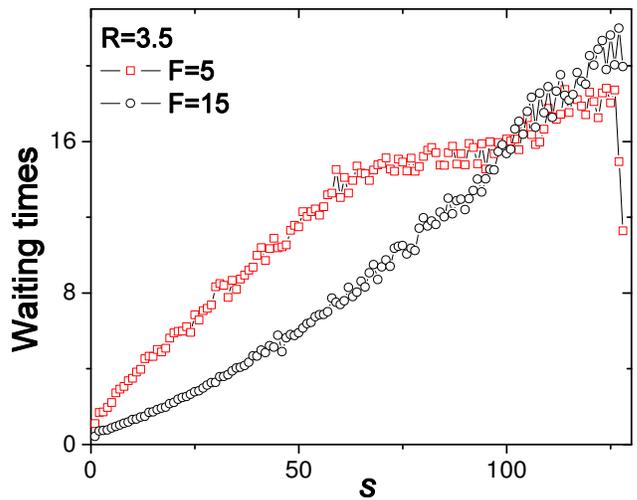}
\caption{(Color online) Waiting time distribution for $N=128$ and $R=3.5$ under different $F$. The waiting time of 
monomer $s$ is defined as the average time between the events that monomer $s$ and monomer $s+1$ exit the pore.
        }
\label{Fig3}
\end{figure}

\begin{figure}
\includegraphics*[width=\figurewidth]{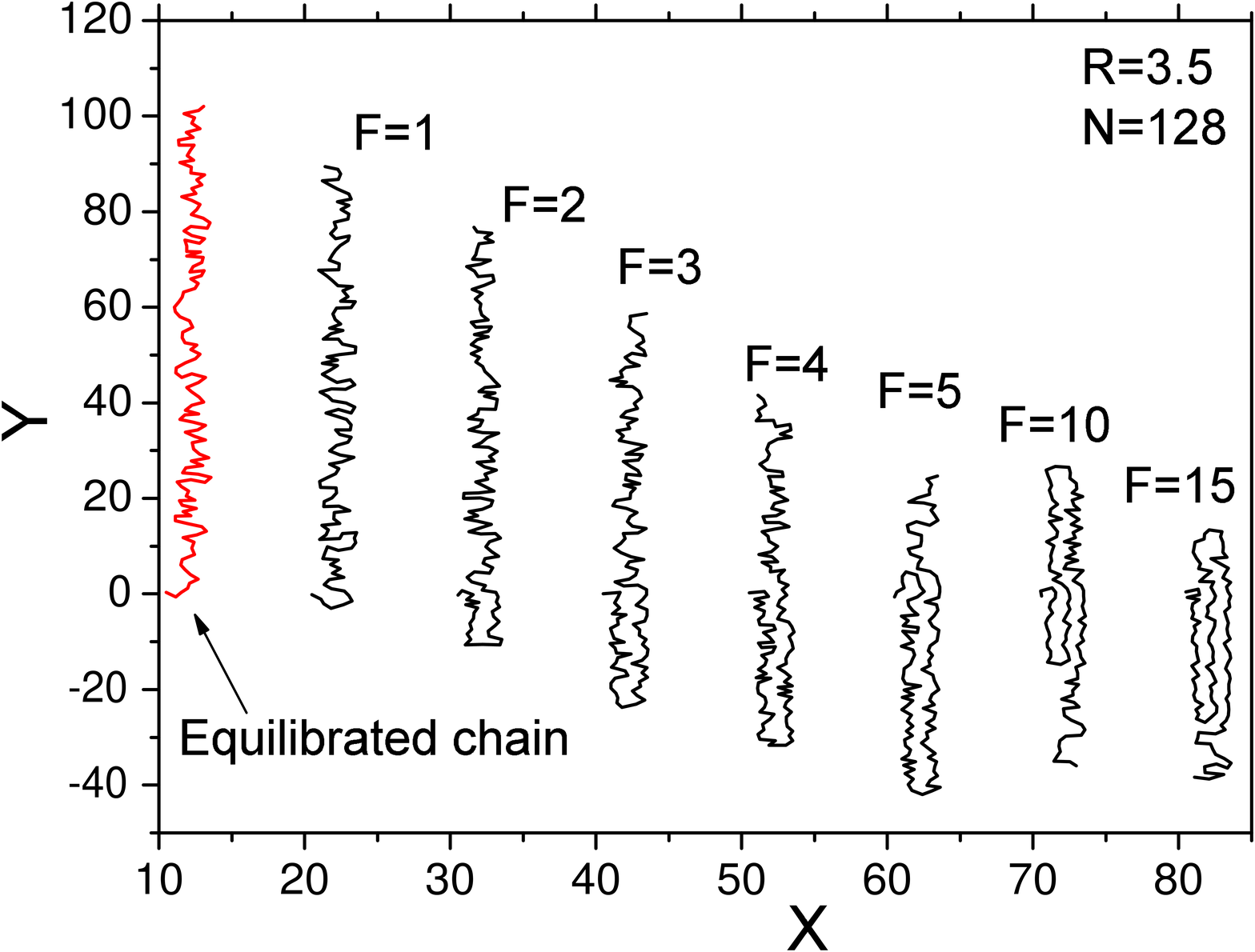}
\caption{(Color online) Typical chain conformation at the moment just after translocation for $N=128$ and
$R=3.5$ under different $F$.
        }
\label{Fig4}
\end{figure}

As shown in Fig. \ref{Fig2} for unconfined system $R=\infty$, the dependence of $\tau$ on the driving force scales 
as $F^{-1}$ for weak driving forces $F\le4$.
This simple scaling behavior can be understood by considering the steady state motion
of the polymer through the nanopore. The average velocity is determined by balancing the
frictional damping force (proportional to the velocity) with the external driving force.
This leads to an average velocity proportional to the driving force $F$, and hence a
translocation time $\tau\sim F^{-1}$.
For $F>4$, $\tau\sim F^{-0.84}$ due to a pronounced non-equilibrium situation where the chain is highly distorted, 
as illustrated in our previous work \cite{Luo8}.
For this case, only part of the chain on the \textit{cis} side can respond immediately, while the remaining
part near the chain end does not feel the force yet. As a result, a part of the chain on the \textit{cis} side is 
deformed to a trumpet and even stem-and-flower shape, while the translocated portion on the \textit{trans} side has 
a compact spherical shape, as it does not have time to diffuse away from the pore exit. In a recent theoretical 
study \cite{Sakaue}, the effect of a trumpet shape of the chain on the \textit{cis} side, on the translocation 
dynamics was found to cause a breakdown of the $\tau\sim F^{-1}$ scaling. However, this theory neglects effects 
due to the compacted chain structure on the
\textit{trans} side.

However, for confined systems the effect of the driving force on the translocation
time is completely different, as shown in Fig. \ref{Fig2} for $R=3.5$. With increasing $F$,
the translocation time initially decreases rapidly, and then almost saturates for $F\ge 10$.
As noted above, $\tau \sim \frac{N^{\alpha}}{F(1-CR^{\gamma}/F)}$ for an equilibrium process.
With decreasing $F$, the resisting force $f(R)$ is more and more important and it slows down
translocation. This is the reason why the initial slope is faster than that for translocation
into a free \textit{trans} side. With increasing $F$, $f(R)$ becomes important again, which indicates
that the resisting force $f(R)$ induced by crowding effects from already translocated monomers
plays an important role in the observed behavior.

To further understand this behavior, we examine the dynamics of a single segment passing
through the pore during translocation. The nonequilibrium nature of translocation has a
significant effect on it. We have numerically calculated the waiting times for all monomers
in a chain of length $N$. We define the waiting time of monomer $s$ as the average time
between the events that monomer $s$ and monomer $s+1$ exit the pore.
In our previous work, we found that the waiting time depends strongly on the monomer positions
in the chain under the driving force $F=5$ \cite{Huopaniemi1,Luo2}. For relatively short polymers,
such as $N=100$, the monomers in the middle of the polymer need the longest time to translocate and
the distribution is close to symmetric.
In contrast, for the confined system with $R=3.5$ and $N=128$, we find
a slow increase after the monomers in the middle of the polymer (i.e.,
beyond $s=N/2$) and a further, more rapid increase
after $s\approx100$ for $F=5$, see Fig. \ref{Fig3}.
Increasing the driving force further to $F=15$, the waiting time rapidly increases throughout.
Moreover, it takes longer for monomers at $s>100$ to exit the pore for strong driving
force $F=15$ than that for $F=5$. These results indicate that during later stages of translocation
the high density of segments in the channel slows down the translocation and the inverse
proportionality $\tau\sim F^{-1}$ breaks down.
Especially for strong driving forces the crowding effect is more pronounced, leading to almost
the same translocation time for $F=10$ and $F=15$.
These results are different from translocation into two parallel walls (3D),  where we observed
a relatively fast turnover between the scaling $\tau \sim F^{-1.25}$ at weak
forces and the behavior $\tau \sim F^{-0.8}$ at strong forces \cite {Luo15}.

Fig. \ref{Fig4} shows the typical chain conformation at the moment just after translocation
for $N=128$ and $R=3.5$ under different $F$.
Compared with the equilibrated chain, the chains become more compressed with increasing $F$.
In particular, a folding of the chain is observed for $F\ge2$, reflecting stronger resisting
forces with increasing $F$.
At $F\ge10$ there even occur triple layers in the chain configuration.
As can be seen the chains in this folded state are almost
entirely devoid of the larger-amplitude undulations observed in the equilibrium
configuration. This also provides a way to induce chain folding by driving
polymers into unidimensionally confined environments such as small channels.

We stop to note that here we confine the analysis to the variation of the
driving force $F$. As demonstrated in Ref.~\cite{Luo8} the variation of the
friction coefficient $\xi$ has a similar effect.

\subsection{Translocation time as a function of the chain length}

Previously \cite{Luo2,Huopaniemi1,Luo3}, we investigated the
polymer translocation into an unconfined \textit{trans} side ($R=\infty$)
under an external driving force $F=5$ in the pore.  Both the 2D fluctuating
bond model with single Monte Carlo moves \cite{Luo2} and Langevin
\cite{Huopaniemi1,Luo3} simulations show $\tau \sim N^{2\nu}\sim N^{1.5}$
for relatively short chains $N \le 200$.

\begin{figure}
\includegraphics*[width=\figurewidth]{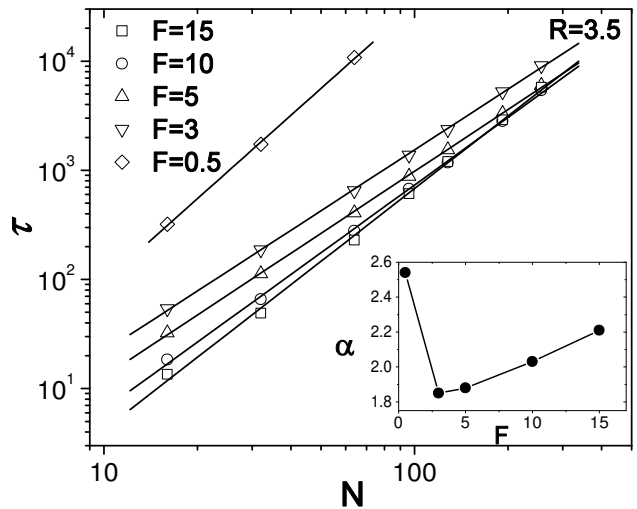}
\caption{Translocation time $\tau$ as a function of the chain length $N$ for
$R=3.5$ and different $F$ in 2D.
The insert shows the scaling exponent $\alpha$ of $\tau$ with
$N$ for different driving forces $F$. For $F$=0.5, 3, 5, 10 and 15,
$\alpha=2.54\pm 0.04$, $1.85\pm 0.01$, $1.88\pm 0.01$, $2.03\pm 0.02$,
and $2.21\pm 0.04$, respectively.
        }
\label{Fig5}
\end{figure}

However, for polymer translocation into the region between two strips, the
dynamics is completely different.
As shown in Fig. \ref{Fig5}, for the same driving force $F=5$ for $R=3.5$, we
find $\tau \sim N^{1.88\pm0.01}$ and translocation velocity
$v \sim N^{1.08\pm0.02}$ close to the inversely linear scaling $v\sim N^{-1}$.
During the translocation process, the chain moves a distance of $R_g+R_{\parallel}$
instead of $2R_g$ for an unconfined system, where $R_{\parallel}$ is the
longitudinal size of the chain. For a polymer confined between two strips
embedded in 2D, the longitudinal size of the chain is $R_{\parallel}\sim N$
(see above scaling results from the blob picture),
which is the dominant term over the second term, $R_g\sim N^{\nu_{\mathrm{2D}}}$.
Thus, the translocation time roughly scales as
$\tau \sim \frac{R_{\parallel}}{v} \sim N^2$.
Decreasing the driving force to $F=3.0$, the scaling exponent almost does not
change, $\tau \sim N^{1.85\pm0.01}$.
Decreasing the driving force further to $F=0.5$, we find the translocation
dynamics crosses over to another regime with $\alpha=2.54\pm 0.04$ at least for
$N \le 64$, which is still lower than the expected scaling exponent 3.0
of the relaxation time as a function of the chain length for a chain in a
narrow channel.
Due to expensive computation, we cannot access
cases with longer chains and/or weaker driving forces.
Increasing the driving force $F$ from 5 to 10 and 15, we find $\tau \sim N^{2.03\pm0.02}$
and $\tau \sim N^{2.21\pm0.04}$, respectively. This is, however, an a priori unexpected
increase of $\alpha$ with increasing $F$, which indicates that the translocation is slowed
down for strong driving forces.
Particularly, for $N=256$ the translocation time almost does not change for $F=5$, 10 and 15,
due to the formation of a folded chain conformation.

\begin{figure}
\includegraphics*[width=\figurewidth]{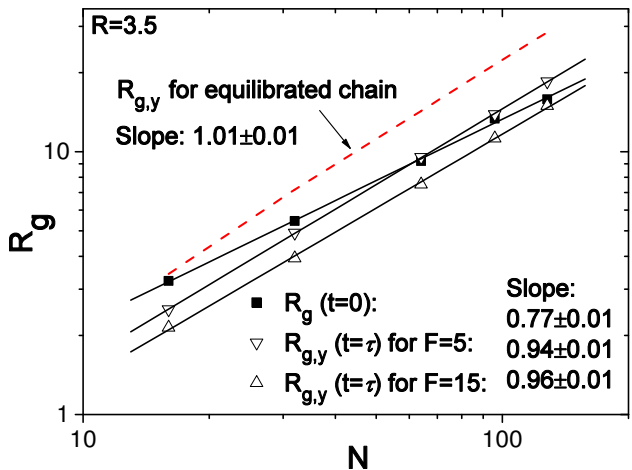}
\caption{(Color online) The radius of gyration of the chain before translocation and at the moment just
after the translocation for $R=3.5$ and different $F$. Here, the $x$ direction is perpendicular
to the wall, and $y$ is the direction along the wall.
        }
\label{Fig6}
\end{figure}

\begin{figure}
\includegraphics*[width=\figurewidth]{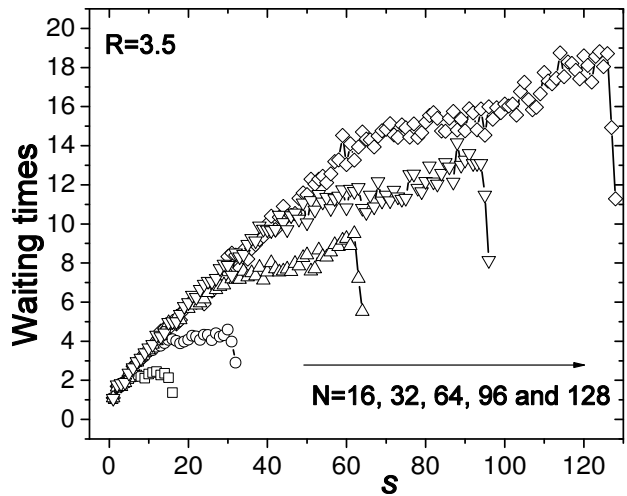}
\caption{Waiting time distribution for different $N$ under the driving force
$F=5$ and $R=3.5$.
        }
\label{Fig7}
\end{figure}

Altogether, compared with an unconfined environment we observe the nonuniversal dependence
of the translocation time $\tau$ on the chain length $N$. Particularly, there is a minimum of
$\alpha$ as a function of $F$, as shown in the insert of the Fig. \ref{Fig5}, due to the more severe
crowding effect for stronger driving forces.
The scaling exponent does not change for $F=15$ compared with $F=5$.
These results are different from the translocation into two parallel walls (3D) with the same $R$,
where for weak driving forces $\tau$ scales exactly in the same manner as the chain relaxation time
and crosses over to exponent 1.37 for strong driving forces \cite {Luo15}.
As intutitively expected, these observations demonstrate that confinement
effects are more relevant in the 1D channel configuration compared to the
2D confinement of the chain sandwiched between two parallel walls.

Fig. \ref{Fig6} shows the radius of gyration of the chain before the translocation and at
the moment just after the translocation for $R=3.5$ and different $F$. Just after the
translocation, the chains are compressed in the $y$ direction compared with the equilibrated chain.
However, the scaling exponent of $R_{g,y}\sim N$ almost does not change even for $F=15$.
Fig. \ref{Fig7} shows the waiting time distribution for different chain lengths
for $F=5.0$ and $R=3.5$. Similar distributions are observed for different chain length.
This behavior reflects the crowding effect of partially
translocated monomers, which greatly slows down translocation.

\subsection{Translocation time as a function of the channel width $R$}

\begin{figure}
\includegraphics*[width=\figurewidth]{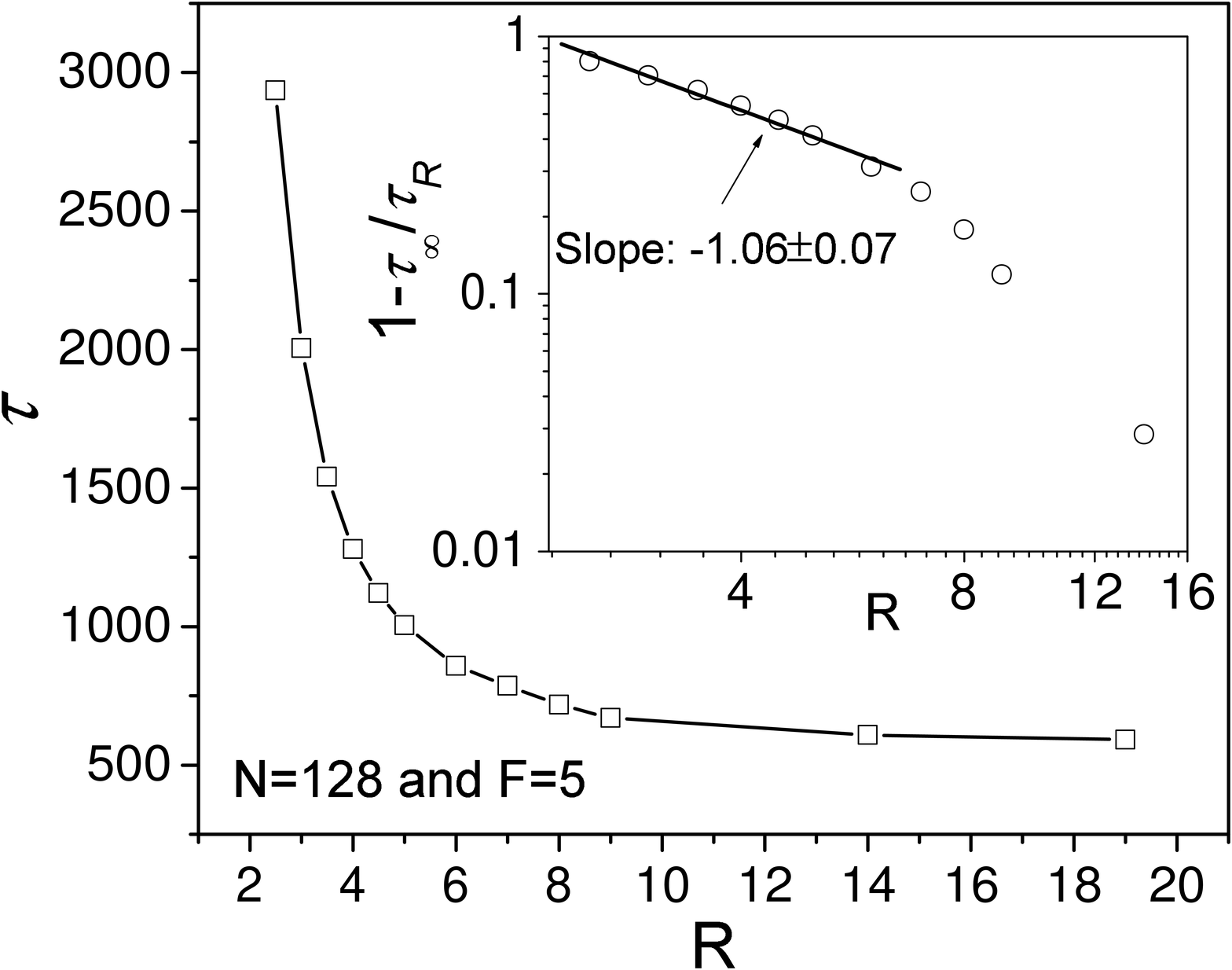}
\caption{Translocation time $\tau$ as a function of $R$ for chain length
$N=128$ under the driving force $F=5$.
        }
\label{Fig8}
\end{figure}

\begin{figure}
\includegraphics*[width=\figurewidth]{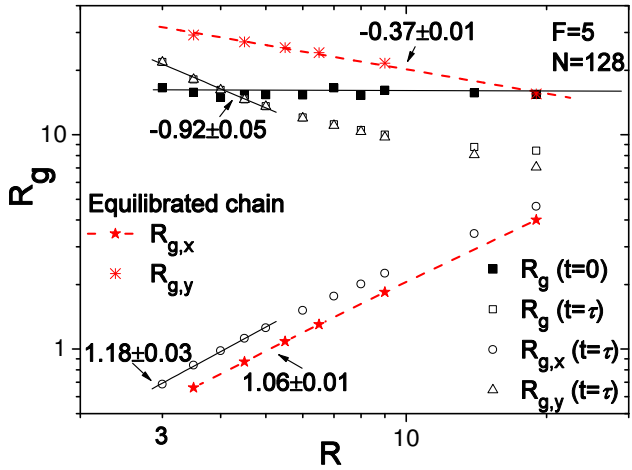}
\caption{(Color online) The radius of gyration of the chain before translocation and at the moment
just after the translocation for $N=128$, $F=5$ versus the channel size $R$. Here, the $x$ direction
is perpendicular to the wall, and $y$ is the direction along the wall.
        }
\label{Fig9}
\end{figure}

\begin{figure}
\includegraphics*[width=\figurewidth]{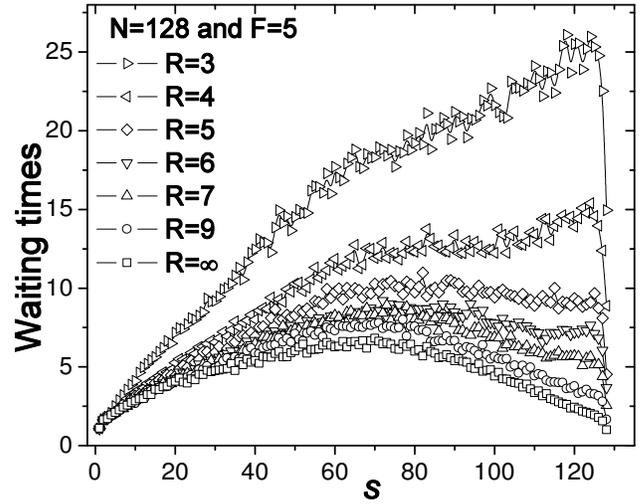}
\caption{Waiting time distribution for different $R$. The chain length $N=128$
and the driving force $F=5$.
        }
\label{Fig10}
\end{figure}

\begin{figure}
\includegraphics*[width=\figurewidth]{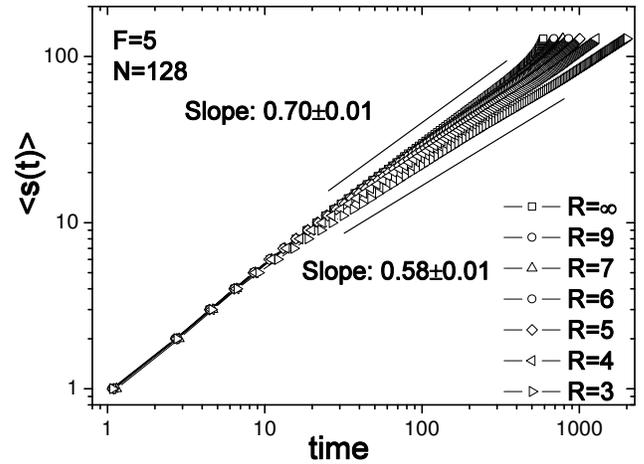}
\caption{The number of translocated monomers as a function of time, $\langle s(t)\rangle$, for
different $R$. The chain length $N=128$ and the driving force $F=5$.
        }
\label{Fig11}
\end{figure}

\begin{figure}
\includegraphics*[width=\figurewidth]{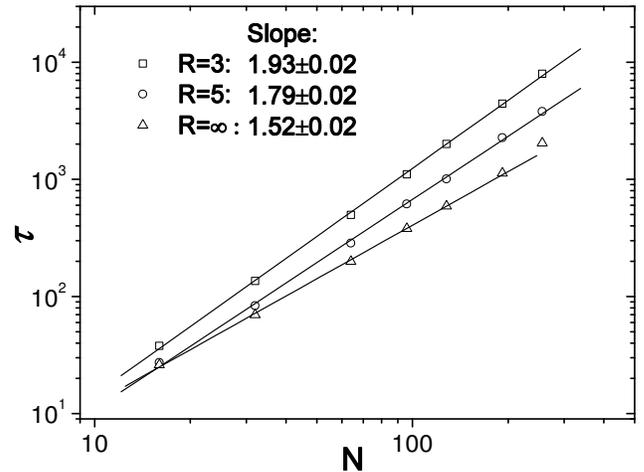}
\caption{Translocation time as a function of the chain length for
different $R$. The driving force $F=5$.
        }
\label{Fig12}
\end{figure}

As shown in Fig.~\ref{Fig8}, there exist two regimes in the behavior of $\tau$
as a function of $R$ for $F=5.0$ and $N=128$. $\tau$ decreases rapidly
with increasing $R$ and then almost saturates for larger $R$. The insert of
Fig. \ref{Fig8} shows $1-\frac {\tau_{\infty}}{\tau}$ as a function of $R$. For
$R \le 6$, we find the exponent $\gamma=-1.06\pm0.07$. For equilibrium translocation process
$f(R)\sim R^{-1/\nu_{2D}}$, which gives the exponent $-1/\nu_{2D}=-1.33$. The
numerical exponent 1.06 does not agree with this result, which indicates that the
translocation is a highly non-equilibrium process.

Fig. \ref{Fig9} shows the radius of gyration of the chain before translocation
and at the moment just after the translocation for $N=128$ and $F=5$ versus the
channel size
$R$. Just after the translocation, the chains are highly compressed along the channel
and $R_{g,y}\sim R^{-0.92}$ for $R\le5$ compared with the
equilibrated chains, where $R_{g,y}\sim R^{-0.37}$ as predicted.

Fig. \ref{Fig10} depicts the waiting time distribution for different $R$. For
larger $R$, we observe a symmetric distribution with respect to the middle
monomer $s=N/2$ \cite{Luo2,Huopaniemi1}. With decreasing $R$, the waiting times
increase. Particularly, it takes much longer time for monomers $s>N/2$ to exit
the pore. For $R=5$, the waiting times approximately saturate after $s>N/2$, while they
continuously increase for $R\le4$. This behavior is due to the crowding of the
translocated monomers.
As to the first moment of the translocation coordinate $\langle s(t)\rangle \sim t^\beta$,
previous results \cite {Luo6} show $\beta\approx 1/\alpha=0.67$ with $\alpha \approx 2\nu=1.50$
for $R=\infty$.
Fig. \ref{Fig11} shows $\langle s(t)\rangle\sim t^\beta$
for $N=128$ and $F=5$ for different $R$.
For $R=\infty$, $\beta=0.70\pm0.01$ as expected and it continuously decreases to $0.58\pm0.01$
for $R=3$. Here, $\beta < 1$ is a signature of anomalous diffusion \cite{Ralf}.
The values of $\beta$ in Fig. \ref{Fig11} imply that the scaling exponent $\alpha$ can change
from $\approx 2\nu$ to $\approx 2$, depending on $R$ due to highly non-equilibrium effects.
This is clearly indicated in Fig. \ref{Fig12}, where $\alpha=1.52\pm0.02$, $1.79\pm0.02$
and $1.93\pm0.02$ for $R=\infty$, 5 and 3, respectively.

\section{Conclusions} \label{chap-conclusions}

Using two-dimensional Langevin dynamics simulations, we investigated the
dynamics of polymer translocation into a narrow channel of width $R$ through
a nanopore under a driving force $F$. Due to the crowding effect induced by
the partially translocated monomers, the translocation dynamics is greatly
changed compared with an unconfined environment. Namely we observe a
nonuniversal dependence of the translocation time $\tau$ on the chain length
$N$: $\tau$ initially decreases rapidly and then saturates with increasing $R$,
and,
moreover, an $R$ dependence of the scaling exponent $\alpha$ in the law $\tau
\sim N^{\alpha}$ is observed. The inversely linear scaling of $\tau$ with $F$
breaks down and we observe a minimum of $\alpha$ as a function of $F$.
These behaviors are clearly interpreted in terms of the waiting time of an
individual segment for passing through the pore during translocation, as well
as in terms of statistical quantities such as the width of the chain. In
particular we observe folded configurations of the chain in the channel,
at high driving forces even a structure with a double-fold exists.
Similar effects of nonequilibrium nature are expected in biological cells.
There, the passage of a translocating chain is opposed by a crowded
environment in which larger biomolecules occupy more than 40\% of the
volume \cite{Review,Packaging}.

Our findings are of interest from a purely polymer physics point of view,
contributing to a more complete picture of polymer translocation through a
narrow pore, once more pronouncing the importance to consider nonequilibrium
situations.
A direct applicatoin of our findings may be of relevance for parallel fluidic
channel setups consisting of stacked parallel slices. In each slice the chain
initially is in an unconfined two-dimensional conformation, before being
forced through the pore into the effectively 1D channel.

\begin{acknowledgments}
K. L. acknowledges the support from the University of Science and Technology
of China through its startup funding of CAS Bairen Program. This work has
been supported in part by the Deutsche Forschungsgemeinschaft (DFG).
\end{acknowledgments}

\end{document}